\documentclass[12pt]{iopart}
\usepackage{iopams}
\usepackage{euscript}
\usepackage{amssymb}
\usepackage{graphicx}
\expandafter\let\csname equation*\endcsname\relax
\expandafter\let\csname endequation*\endcsname\relax
\usepackage{amsmath}
\usepackage{amsbsy}
\usepackage{amsthm}
\usepackage{bbm}
\usepackage{bm}
\usepackage{epsfig}
\usepackage{epstopdf}
\usepackage{dsfont}
\usepackage{color}
\usepackage[colorlinks]{hyperref}
\usepackage[figure,table]{hypcap}
\usepackage[symbol]{footmisc}
\usepackage{enumerate}
\usepackage{geometry}
\hypersetup{
	bookmarksnumbered,
	pdfstartview={FitH},
	citecolor={darkgreen},
	linkcolor={darkblue},
	urlcolor={darkblue},
	pdfpagemode={UseOutlines}}
\definecolor{darkgreen}{RGB}{50,190,50}
\definecolor{darkblue}{RGB}{0,0,190}
\definecolor{darkred}{RGB}{238,0,0}
\usepackage{soul}

\newcommand{\bra}[1]{\ensuremath{\left\langle\right. #1 \left.\right|}}
\newcommand{\ket}[1]{\ensuremath{\left|\right. #1 \left.\right\rangle}}

\newcommand{\comm}[2]{\ensuremath{\left[\right.\! #1 \,, #2 \!\left.\right]}}

\DeclareMathOperator{\diag}{diag}

\begin{document}

\title{On the robustness of entanglement in analogue gravity systems}

\date{January 2013}
\author{D E Bruschi$^{1,2}$\footnote{Present address: Racah Institute of Physics and Quantum Information Science,
Givat Ram, Hebrew University of Jerusalem, 91904, Jerusalem, Israel}, N Friis$^{2,3}$, I Fuentes$^{2}$\footnote{Ivette Fuentes previously published as Ivette Fuentes-Guridi and Ivette Fuentes-Schuller} and S Weinfurtner$^{2,4}$}
%
%

\address{
$^{1}$School of Electronic and Electrical Engineering,
University of Leeds,
Woodhouse Lane,
Leeds LS2 9JT,
United Kingdom}
\address{
$^{2}$School of Mathematical Sciences,
University of Nottingham,
University Park,
Nottingham NG7 2RD,
United Kingdom}
\address{
$^{3}$Institute for Quantum Optics and Quantum Information,
Austrian Academy of Sciences,
Technikerstr. 21a,
A-6020 Innsbruck,
Austria}
\address{
$^{4}$SISSA,
Via Bonomea 265, 34136,
Trieste, Italy and
INFN, Sezione di Trieste}
\eads{\href{mailto:david.edward.bruschi@gmail.com}{david.edward.bruschi@gmail.com} and \href{mailto:nicolai.friis@uibk.ac.at}{nicolai.friis@uibk.ac.at}}

\begin{abstract}
We investigate the possibility to generate quantum-correlated quasi-particles utilizing analogue gravity systems.
The quantumness of these correlations is a key aspect of analogue gravity effects and their presence
allows for a clear separation between classical and quantum analogue gravity effects. However, experiments in analogue
systems, such as Bose-Einstein condensates, and shallow water waves, are always conducted at non-ideal conditions,
in particular, one is dealing with dispersive media at nonzero temperatures. We analyze the influence of the initial temperature on the entanglement generation in analogue gravity phenomena. We lay out all the necessary
steps to calculate the entanglement generated between quasi-particle modes and we analytically derive an upper bound on the maximal temperature at which given modes can still be entangled. We further investigate a mechanism to enhance the quantum correlations. As a particular example we analyze the robustness of the entanglement creation against thermal noise in a sudden quench of an ideally homogeneous Bose-Einstein condensate, taking into account the super-sonic dispersion relations.
\end{abstract}
\date{\today}
\begin{indented}
\item New Journal of Physics \textbf{15} (2013) 113016\\
doi:\href{http://dx.doi.org/10.1088/1367-2630/15/11/113016}{10.1088/1367-2630/15/11/113016}
\end{indented}



\maketitle

\section{Introduction}\label{sec:introduction}
%

\emph{Can quantum effects in curved spacetimes be simulated in compact, laboratory-based experimental setups?} Following the
formal analogy between quantum field theory on curved spacetimes and classical fluid systems that was established by W.~G.~Unruh
in~\cite{Unruh1981}, this question has captivated researchers for decades (see, e.g., Ref.~\cite{BarceloLiberatiVisser2005} for
a recent review). In particular, the prospect of accessible experimental setups to test the quantum effects of varying spacetime
backgrounds has motivated scientists, who have subsequently directed their ingenuity and effort towards the study of analogue
gravity systems. The range of physical systems in which such simulations can be performed is vast, reaching from actual shallow
water waves~\cite{RousseauxMathisMaissaPhilbinLeonhardt2008,WeinfurtnerTedfordPenriceUnruhLawrence2011,
JannesPiquetMaissaMathisRousseaux2011,WeinfurtnerTedfordPenriceUnruhLawrence2013}, and Bose-Einstein
condensates (BECs)~\cite{GarayAnglinCiracZoller2000,LahavItahBlumkinRinottZayatsSteinhauer2010,JaskulaPartridgeBonneauRuaudelBoironWestbrook2012}, to laser pulse
filaments~\cite{Belgiorno-Faccio2010,Rubino-Faccio2012}, to name but a few.

A central aim in such studies is the observation of radiation that can be associated
to quantum pair creation processes, for instance, to the Hawking-, Unruh- and the dynamical Casimir effect. All of these effects rely on similar mechanisms in quantum field theory, i.e., particle creation due to time-dependent gravitational fields and boundary conditions, or the presence of horizons. It is then only natural to ask what are the criteria for associating the effects of quantum field theory with the analogue systems. How can these criteria \emph{distinguish quantum from classical scattering processes?}
In particular, since it is a matter of ongoing debate which systems actually exhibit quantum effects, while others, e.g., water waves, are not expected to produce genuine quantum effects at all. One aspect that has already been studied theoretically~\cite{RecatiPavloffCarusotto2009}, and has been rigourously tested in experiments is the spectrum of the radiation and its relation to the effective gravitational
field (see, e.g., Ref.~\cite{WeinfurtnerTedfordPenriceUnruhLawrence2011}). However, it might be argued that the ingredient that is missing so far is the verification of the quantumness of the observed radiation via a direct detection of entanglement, which cannot be inferred from the spectrum alone without additional information. Efforts have
been directed towards addressing this issue by studying non-classical behaviour as captured by sub-Poiss{\'o}nian statistics and the
connected violation of Cauchy-Schwarz inequalities~\cite{Kheruntsyan-Westbrook2012,deNovaSolsZapata2012}. However, the most paradigmatic
quantum mechanical feature\textemdash \emph{entanglement}\textemdash  has not yet been verified in analogue gravity systems, despite the fact that typical pair
creation processes in curved spacetime scenarios~\cite{BirrellDavies:QFbook} involve the creation of mode entanglement.
For instance, entanglement is created between modes of quantum fields in an expanding spacetime (see, e.g.,
Ref.~\cite{BallFuentes-SchullerSchuller2006}), a system that has been previously studied in the context of analogue gravity (see, for example, Refs.~\cite{BarceloLiberatiVisser2003,CalzettaHu2003,FedichevFischer2003}).
However, the presence of quantum correlations depends not only on the presence of entangling processes, but also on the initial state of the system. The temperature in the system needs to be low enough to allow entanglement to be generated.

Here we study how robust possible entanglement generation phenomena in analogue gravity systems, e.g., in BEC simulations of an expanding universe~\cite{JainWeinfurtnerVisserGardiner2007,WeinfurtnerWhiteVisser2007,BallFuentes-SchullerSchuller2006} and the dynamical
Casimir effect~\cite{JaskulaPartridgeBonneauRuaudelBoironWestbrook2012}, are against such thermal noise. For this purpose we connect the techniques for quantum information processing with continuous variables (see, e.g., Ref.~\cite{AdessoIlluminati2005}) and relativistic quantum information (RQI). During the last decade the RQI community has developed techniques to quantify entanglement in quantum field theory (for a review see, e.g., Ref.~\cite{AlsingFuentes2012}). While previous work, see, e.g., Refs.~\cite{HorstmannReznikFagnocchiCirac2010,HorstmannSchuetzholdReznikFagnocchiCirac2010}, has already provided model specific discussions, in this article we apply the framework recently developed within RQI~\cite{FriisFuentes2013,BruschiDraganLeeFuentesLouko2012} to establish the first general description of entanglement generation in analogue gravity systems, including such effects as initial temperatures
and nonlinear dispersion relations. Connecting RQI and analogue gravity promises to be a fruitful endeavour. We address the central question: \emph{Is it in principle possible to observe quantum correlations in analogue gravity systems?} Naively, the answer is: Yes. But, the effects are highly sensitive to the levels of thermal noise. We provide closed analytical expressions for the required maximal background temperatures. Our framework applies directly to all analogue gravity systems where only two degrees of freedom are mixed. For example, all homogeneous analogue gravity systems and such inhomogeneous systems where only two phononic modes become entangled. In cases where more degrees of freedom are mixed, for example in the presence of inhomogeneities, our results can be considered as a limiting case that supplies an upper bound for the allowed temperature by virtue of the monogamy of entanglement~\cite{HiroshimaAdessoIlluminati2007}. Our criteria thus represent critical benchmarks that need to be taken into account by the next generation of analogue gravity experiments. In addition, we consider a scheme\textemdash  entanglement resonances\textemdash to enhance the generation of quantum correlations to overcome temperature restrictions. We illustrate our results for a particular example, an ideally homogeneous BEC undergoing a single sudden quench, i.e., a change of the speed of sound, or series of such transformations, for which we explicitly include the effects of nonlinear dispersion and we assume that the dispersive effects are negligible.

\section{Entanglement in analogue gravity systems}\label{sec:Entanglement in analogue gravity systems}

Let us start with a brief overview of the description and quantification of entanglement. For an introductory
review see, e.g., Ref.~\cite{Bruss2002}. A quantum state is called \emph{separable} with respect to the
bipartition into subsystems A and B if the corresponding density operator $\rho_{\raisebox{0.0pt}{\tiny{AB}}}$
can be written as
\begin{equation}
\rho_{\raisebox{0.0pt}{\tiny{AB}}}=\sum_{i}p_{i}
        \ket{\!\psi^{\raisebox{0.0pt}{\tiny{A}}}_{i}\!}\!\bra{\!\psi^{\raisebox{0.0pt}{\tiny{A}}}_{i}\!}\otimes
        \ket{\!\psi^{\raisebox{0.0pt}{\tiny{B}}}_{i}\!}\!\bra{\!\psi^{\raisebox{0.0pt}{\tiny{B}}}_{i}\!},
\end{equation}
where $\sum_{i}p_{i}=1$, $p_{i}\geq0$, and $\ket{\psi^{\raisebox{0.0pt}{\tiny{A/B}}}_{i}}$ are pure states of the subsystems
A or B respectively. Such states can contain classical correlations, i.e., correlations that depend on the
local (in the sense of the subsystems A and B) choice of basis. Quantum states that cannot be decomposed in
the form of $\rho_{\raisebox{0.0pt}{\tiny{AB}}}$ above are called \emph{entangled}. A formal way to quantify the entanglement of a given state
$\rho_{\raisebox{0.0pt}{\tiny{AB}}}$ is the \emph{entanglement of formation} $E_{oF}$, defined as~\cite{BennettDiVincenzoSmolinWootters1996}
\vspace*{-2mm}
\begin{align}
    E_{oF}(\rho_{\raisebox{0.0pt}{\tiny{AB}}})  &=\,\inf_{\{p_{i},\psi^{\raisebox{0.0pt}{\tiny{AB}}}_{i}\}}
        \sum\limits_{i}p_{i}\,\mathcal{E}(\ket{\!\psi^{\raisebox{0.0pt}{\tiny{AB}}}_{i}\!})\,.
    \label{eq:EoF definition}
\end{align}
Here $\mathcal{E}(\ket{\!\psi^{\raisebox{0.0pt}{\tiny{AB}}}\!})$ is the entropy of entanglement, given by the von~Neumann entropy of the reduced state $\rho_{\raisebox{0.0pt}{\tiny{A}}}=\tr_{\raisebox{0.0pt}{\tiny{B}}}
\ket{\!\psi^{\raisebox{0.0pt}{\tiny{AB}}}\!}\!\bra{\!\psi^{\raisebox{0.0pt}{\tiny{AB}}}\!}$. The minimum in Eq.~(\ref{eq:EoF definition}) is taken over all pure state decompositions $\{p_{i},\psi^{\raisebox{0.0pt}{\tiny{AB}}}_{i}\}$ that realize the state $\rho_{\raisebox{0.0pt}{\tiny{AB}}}$, i.e., such that $\rho_{\raisebox{0.0pt}{\tiny{AB}}}=\sum_{i}p_{i}\ket{\!\psi^{\raisebox{0.0pt}{\tiny{AB}}}_{i}\!}\!\bra{\!\psi^{\raisebox{0.0pt}{\tiny{AB}}}_{i}\!}$.
For mixed states of arbitrary dimension this measure is not computable but for some special cases, including those we discuss
in this paper, the minimization in Eq.~(\ref{eq:EoF definition}) can be carried out and $E_{oF}$ can be computed analytically.

In the analogue gravity context that we want to consider here, the subsystems $A$ and $B$ will be two (out of an ensemble of possibly infinitely many) bosonic modes, e.g., associated to phonons in a BEC~\cite{{WeinfurtnerWhiteVisser2007}}. The corresponding annihilation and creation operators, $a_{i}$ and $a^{\dagger}_{i}$ satisfy the commutation relations $\comm{a_{i}}{a^{\dagger}_{j}}=\delta_{ij}$ and $\comm{a_{i}}{a_{j}}=0$. The operators $a_{i}$ and $a_{i}^{\dagger}$ may be combined into
the quadrature operators $q_{j}:=\frac{1}{\sqrt{2}}(a_{j}+a^{\dagger}_{j})$ and $p_{j}:=\frac{-i}{\sqrt{2}}(a_{j}-a^{\dagger}_{j})$, which, in turn,
can be collected into the vector
\begin{align}
  \mathbb{X}  &:=\,\bigl(q_{1},p_{1},q_{2},p_{2},\ldots\bigr)^{\mathrm{T}}\,.
    \label{eq:vector of first moments}
\end{align}
For the particularly important class of \emph{Gaussian states}, which includes, e.g., the vacuum and thermal states, all the information
about the state $\rho$ is encoded in the first moments $\left\langle\mathbb{X}_{i}\right\rangle_{\!\rho}$ and the second moments
\begin{align}
    \Gamma_{ij}:=\bigl<\mathbb{X}_{i}\mathbb{X}_{j}+\mathbb{X}_{j}\mathbb{X}_{i}\bigr>_{\!\rho}
    -2\bigl<\mathbb{X}_{i}\bigr>_{\!\rho}\bigl<\mathbb{X}_{j}\bigr>_{\!\rho}\,,
    \label{eq:covariance matrix}
\end{align}
where $\bigl<\mathbb{X}_{i}\bigr>_{\!\rho}$ denotes the expectation value of the operator $\mathbb{X}_{i}$ in the state $\rho$ (see
Ref.~\cite{AdessoIlluminati2005}). Furthermore, the \emph{covariance matrix} $\Gamma$ contains all the relevant information about the
entanglement between the modes.

Let us now consider a typical transformation occurring in analogue gravity systems, for example, the generation of phonons in
a BEC that is undergoing a sudden change of the speed of sound. Such transformations are described by \emph{Bogoliubov transformations}, i.e.,
linear transformations
\begin{align}
    \tilde{a}_{m}   &=\,\sum\limits_{n}\bigl(\alpha^{*}_{mn}a_{n}\,-\,\beta^{*}_{mn}a^{\dagger}_{n}\bigr)
    \label{eq:Bogoliubov transformation}
\end{align}
between two sets of annihilation and creation operators, $\{(a_{i},a^{\dagger}_{i})\}$ and
$\{(\tilde{a}_{i},\tilde{a}^{\dagger}_{i})\}$, $(i=1,2,\ldots)$, that leave the canonical commutation relations invariant. The
Bogoliubov transformation induces a unitary transformation on the Hilbert space of states. In phase space, on the other hand,
these unitaries are realized as \emph{symplectic} transformations $S$ that satisfy $S\Omega S^{\mathrm{T}}=\Omega$, where the symplectic
form $\Omega$ is defined by the relation $i\Omega_{mn}=\bigl[\mathbb{X}_{m},\mathbb{X}_{n}\bigr]$. The transformation~$S$ can be
expressed in terms of the coefficients $\alpha_{mn}$ and $\beta_{mn}$, which allows us to quantify the entanglement that is being
generated between the modes for a wide variety of scenarios (see Ref.~\cite{FriisFuentes2013}).

At this point it is crucial to notice that, in the most favorable scenarios in analogue gravity (e.g., homogenous systems) the structure of the Bogoliubov transformations can be particularly simple. In these cases, the transformations mix pairs of degrees of freedom $p,p^{\prime}$ and the only non vanishing Bogoliubov coefficients are $\alpha_{pp},\alpha_{p^{\prime}p^{\prime}}$ and $\beta_{pp^{\prime}}$. For example, in the case of homogeneous systems, the absence of boundaries that couple counter-propagating modes along with momentum conservation implies that only coefficients $\alpha_{kk}$ and $\beta_{k(-k)}\ (\forall k)$ are non-zero, where $k$ labels the momentum. This occurs even if the dispersion relation allows for more than two modes to correspond to the same frequency. In other words, the transformation cannot shift the momenta of individual excitations but it allows for the creation of (quasi-)particle pairs with equal but opposite momenta. Furthermore, in systems that are inherently inhomogeneous, such as analogue black hole setups, the coupling can be limited to pairs of frequencies due to the system's stationarity. In order to illustrate our techniques we specialize to systems where the mixing occurs between modes of opposite momenta. This simple structure of the Bogoliubov coefficients permits us to consider the covariance matrix for any pairs of modes~$k$ and~$-k$ independently of any other modes of the continuum. In the following we consider this simple structure to be an approximation for inhomogeneous systems with minor density fluctuations. In realistic setups, entanglement is generated between more than two degrees of freedom. For example, in the setups we address here, the effect of the inhomogeneity is to distribute the entanglement that is generated also across modes with different momenta~\cite{RichartzWeinfurtnerPennerUnruh2009}, introducing additional noise in the reduced state of the modes $k$ and $-k$. Due to monogamy restraints~\cite{HiroshimaAdessoIlluminati2007} this will decrease the amount of entanglement produced between these modes, such that the results we obtain can be considered as upper bounds on the entanglement generation.

Dispersive effects on the other hand, e.g., in effective Friedmann-Robertson-Walker spacetimes, are easily taken into account in terms of modified Bogoliubov coefficients. These can be obtained by solving the corresponding equation of motion for the field modes, involving forth order derivatives in space for sub- and super-luminal dispersion relations. For the two chosen modes the symplectic transformation $S$ can be written as
\begin{align}
    S   &=\,\begin{pmatrix}
        \mathcal{M}_{kk}    &   \mathcal{M}_{k(-k)} \\
        \mathcal{M}_{(-k)k} &   \mathcal{M}_{(-k)(-k)}
    \end{pmatrix}\,,
    \label{eq:symplectic transformation for two modes}
\end{align}

where the $2\times2$ blocks are given by
\begin{subequations}
\label{eq:symplectic transformation alpha and beta blocks}
\begin{align}
    \mathcal{M}_{nn}   &=\,\begin{pmatrix}
        \mathrm{Re}(\alpha_{nn})    &   \mathrm{Im}(\alpha_{nn})    \\
        -\mathrm{Im}(\alpha_{nn})   &   \mathrm{Re}(\alpha_{nn})
    \end{pmatrix}\,,\
    \label{eq:symplectic transformation alpha block}\\[1.5mm]
    \mathcal{M}_{n(-n)}   &=\,\begin{pmatrix}
        -\mathrm{Re}(\beta_{n(-n)})     &   \mathrm{Im}(\beta_{n(-n)})      \\
        \mathrm{Im}(\beta_{n(-n)})      &   \mathrm{Re}(\beta_{n(-n)})
    \end{pmatrix}\,,
    \label{eq:symplectic transformation beta block}
\end{align}
\end{subequations}

with $n=k,-k$. The unitarity of the transformation implies
\begin{align}
    |\alpha_{kk}|^{2}\,-\,|\beta_{k(-k)}|^{2}   &=\,1\,,
    \label{eq:bogo condition I}
\end{align}
and, in general for $\Theta\in\mathbb{R}$, we have
\begin{subequations}
\label{eq:bogo conditions II}
\begin{align}
    \alpha_{kk} &=\,e^{i\Theta}\alpha_{(-k)(-k)}\,,
    \label{eq:bogo conditions II alphas}\\[1mm]
    \beta_{k(-k)}   &=\,e^{i\Theta}\beta_{(-k)k}\,,
    \label{eq:bogo conditions II betas}
\end{align}
\end{subequations}
where the phase $\Theta\in\mathbb{R}$ is left undetermined by the unitarity of the transformation. Let us now consider the effect of the Bogoliubov transformation on the
entanglement between the modes $k$ and $-k$, including dispersive and finite temperature effects.
Ideally, these modes are initially in the ground state. However, in every analogue gravity setup, the background temperature is non-zero
(see, e.g., Ref.~\cite{JaskulaPartridgeBonneauRuaudelBoironWestbrook2012}).
We are therefore applying the transformation $S$ to the covariance matrix $\Gamma_{\mathrm{th}}(T)$ of
a thermal state at temperature $T$. Since in our case the modes $k$ and $-k$ have the same initial frequency
$\omega_{\mathrm{in}}=\omega_{\mathrm{in}}(|k|)$ their thermal covariance matrix is proportional to the
identity and given by (see~\cite{AspachsCalsamigliaMunoz-TapiaBagan2009}),
\begin{align}
    \Gamma_{\mathrm{th}}(T) &=\,\coth\bigl(\frac{\hbar\,\omega_{\mathrm{in}}}{2\,k_{\mbox{\tiny{B}}}\,T}\bigr)\mathds{1}\,,
    \label{eq:thermal covariance matrix}
\end{align}
such that the average particle number is distributed according to Bose-Einstein statistics. The transformed state has the form
\vspace*{-2mm}
\begin{small}
\begin{align}
    \tilde{\Gamma}  &=\,S\,\Gamma_{\mathrm{th}}(T)\,S^{\mathrm{T}}\,=\,
    \begin{pmatrix}
        \tilde{\Gamma}_{k}       &   \!C         \\[0.5mm]
        \,C^{\,\mathrm{T}}   &   \,\tilde{\Gamma}_{\!-k}
    \end{pmatrix}\,,
    \label{eq:transformed covariance matrix}
\end{align}
\end{small}
where $C$ is composed of the $2\times2$ matrices of Eq.~(\ref{eq:symplectic transformation alpha and beta blocks}) and the reduced state covariance matrices of the individual modes, $\tilde{\Gamma}_{k}$ and $\tilde{\Gamma}_{\!-k}$, are identical thermal
states
\vspace*{-3mm}
\begin{align}
    \tilde{\Gamma}_{k}  &=\,\tilde{\Gamma}_{\!-k}\,=\,\coth\bigl(\frac{\hbar\,\omega_{\mathrm{in}}}{2\,k_{\mbox{\tiny{B}}}\,T}\bigr)
    (2|\beta_{k(-k)}|^{2}\,+\,1)\,\mathds{1}\,
    \label{eq:single mode transformed covariance matrices}
\end{align}
with non-zero temperature even when the initial temperature $T$ is vanishing. We shall use this fact to define a characteristic
temperature $T_{\mbox{\tiny{E}}}$ of the individual modes via the relation
\begin{align}
    \coth\bigl(\frac{\hbar\,\omega_{\mathrm{out}}}{2\,k_{\mbox{\tiny{B}}}\,T_{\mbox{\tiny{E}}}}\bigr)  &=\,2|\beta_{k(-k)}|^{2}\,+\,1\,,
    \label{eq:entanglement temperature}
\end{align}
where we have taken into account a possible change in frequency, $\omega_{\mathrm{in}}\rightarrow\omega_{\mathrm{out}}$, for fixed
$k$, due to nonlinear dispersion. This \emph{entanglement temperature} $T_{\mbox{\tiny{E}}}$, which corresponds to the Hawking
temperature for a black hole evaporation process, can be attributed purely to the entanglement that is generated from the initial vacuum in a homogeneous system, in complete
analogy to the mixedness that is quantified by the entropy of entanglement in Eq.~(\ref{eq:EoF definition}).

If the first moments $\bigl<\mathbb{X}_{i}\bigr>_{\!\rho}$ of the initial state vanish,
the average particle number after the transformation can be computed from
\begin{align}
\tilde{N}_{k}=\bigl<a_{k}^{\dagger}a_{k}\bigr>=\tfrac{1}{4}\bigl(\tilde{\Gamma}_{11}+\tilde{\Gamma}_{22}-2\bigr),
\end{align}
where the $\tilde{\Gamma}_{ij}$ are the elements of the $4\times4$ covariance matrix $\tilde{\Gamma}$ [see Eq.~(\ref{eq:covariance matrix})], and we obtain
\begin{align}
    \tilde{N}_{k}   &=\,\frac{1}{2}\Bigl(\coth\bigl(\frac{\hbar\,\omega_{\mathrm{out}}}{2\,k_{\mbox{\tiny{B}}}\,T_{\mbox{\tiny{E}}}}\bigr)
                        \coth\bigl(\frac{\hbar\,\omega_{\mathrm{in}}}{2\,k_{\mbox{\tiny{B}}}\,T}\bigr)-1\Bigr)\,,
    \label{eq:average particle number}
\end{align}
which reduces to the usual Bose-Einstein statistics for $\beta_{k(-k)}=0$, while it takes the familiar form~\cite{BirrellDavies:QFbook} \mbox{$\tilde{N}_{k}(T=0)=|\beta_{k(-k)}|^{2}$} starting from the initial vacuum. Initially, i.e., for $T_{\mbox{\tiny{E}}}=0$, the distribution~(\ref{eq:average particle number}) is thermal but after the transformation this is not necessarily the case since $T_{\mbox{\tiny{E}}}=T_{\mbox{\tiny{E}}}(k)\,$. From Eq.~(\ref{eq:single mode transformed covariance matrices}) we can further infer that $\tilde{\Gamma}$ is a \emph{symmetric} state, i.e., $\det\Gamma_{k}=\det\Gamma_{\!-k}$, for which the entanglement of formation $E_{oF}$ can be computed explicitly (see Ref.~\cite{AdessoIlluminati2005}). It is simply given as a function of the parameter $\nu_{\!\small{-}}\geq0$, the smallest eigenvalue of $|i\Omega\,P_{k}\,\tilde{\Gamma}\,P_{k}|$, where $P_{k}=\diag\{1,-1,1,1\}$ represents partial transposition of mode $k$. If $0\leq\nu_{\!\small{-}}<1$ the transformed state $\tilde{\Gamma}$ is entangled and $E_{oF}$ is a monotonously decreasing function of $\nu_{\!\small{-}}$, given by
\begin{align}
    E_{oF}  &=\,\Bigl\{\,  \begin{matrix}  h(\nu_{\!\small{-}})  &   \ \ \mbox{if}\ \ 0\,\leq\,\nu_{\!\small{-}}\,<\,1\\[0.5mm]
    0   &   \ \ \mbox{if}\ \ \nu_{\!\small{-}}\geq\,1\end{matrix}\,,
    \label{eq:EoF for symmetric Gaussian states}
\end{align}
\vspace*{-5mm}
where\\
\begin{small}
\begin{align}
h(x)    &=\,
\frac{(1+x)^{2}}{4\,x}
\ln\frac{(1+x)^{2}}{4\,x}\,-\,
\frac{(1-x)^{2}}{4\,x}
\ln\frac{(1-x)^{2}}{4\,x}\,.
    \label{eq:entropy measure}
\end{align}
\end{small}
For the state $\tilde{\Gamma}$ of Eq.~(\ref{eq:transformed covariance matrix}) we find
\begin{align}
\nu_{\!\small{-}}(T)   &=\,\coth\bigl(\frac{\hbar\,\omega_{\mathrm{in}}}{2\,k_{\mbox{\tiny{B}}}\,T}\bigr)
    (|\alpha_{kk}|-|\beta_{k(-k)}|)^{2}\,.
    \label{eq:nu minus}
\end{align}
While Eqs.~(\ref{eq:EoF for symmetric Gaussian states})-(\ref{eq:nu minus}) completely quantify the entanglement
that is generated by the Bogoliubov transformation in the initial thermal state of temperature $T$, it is
Eq.~(\ref{eq:nu minus}) alone that is needed to determine whether or not any entanglement is created at all. In
particular, we can identify the \emph{sudden death temperature} $T_{\mbox{\tiny{SD}}}$, i.e., the initial temperature
at which a given Bogoliubov transformation no longer generates any entanglement between the fixed modes $k$ and $-k$.
It is determined by the condition $\nu_{\!\small{-}}(T_{\mbox{\tiny{SD}}})=1$. By combining
the entanglement temperature and Eq.~(\ref{eq:nu minus}) we can express this condition as
\vspace*{-2mm}
\begin{align}
    \coth\bigl(\frac{\hbar\,\omega_{\mathrm{out}}}{2\,k_{\mbox{\tiny{B}}}\,T_{\mbox{\tiny{E}}}}\bigr)
    &=\,\coth\bigl(\frac{\hbar\,\omega_{\mathrm{in}}}{k_{\mbox{\tiny{B}}}\,T_{\mbox{\tiny{SD}}}}\bigr)\,,
    \label{eq:sudden death condition}
\end{align}
which, in turn, implies 
\begin{align}
    T_{\mbox{\tiny{SD}}} &=\,2\,\frac{\omega_{\mathrm{in}}}{\raisebox{1.0pt}{$\omega_{\mathrm{out}}$}}\,T_{\mbox{\tiny{E}}}\,.
    \label{eq:sudden death temperature}
\end{align}
This simple relation provides us with a powerful tool to determine if a particular analogue gravity setup can in principle be expected to produce entanglement, for instance in a BEC simulating an expanding universe~\cite{BallFuentes-SchullerSchuller2006}. However, we can also formulate a simple procedure to identify transformations whose repetitions\textemdash  should they be implementable\textemdash  will resonantly enhance the entanglement produced between particular modes. In the following we will study the symplectic representation $S$ of such a repeatable transformation and identify the conditions for enhancing the ability of detecting entanglement.

\section{Entanglement resonances}\label{sec:resonances}

Any symplectic transformation of two modes can be decomposed into a passive transformation $S_{\raisebox{0.0pt}{\tiny{P}}}$, with
$S_{\raisebox{0.0pt}{\tiny{P}}}^{\,T}S_{\raisebox{0.0pt}{\tiny{P}}}=\mathds{1}$, representing rotations and beam splitting, and an active transformation
$S_{\raisebox{0.0pt}{\tiny{A}}}=S_{\raisebox{0.0pt}{\tiny{A}}}^{\,T}$, consisting of single- and two-mode squeezing, i.e., $S=S_{\raisebox{0.0pt}{\tiny{P}}}S_{\raisebox{0.0pt}{\tiny{A}}}$ (see Ref.~\cite{ArvindDuttaMukundaSimon1995}).
From the reduced states in Eq.~(\ref{eq:single mode transformed covariance matrices}) we can easily see that the
transformation described by Eq.~(\ref{eq:symplectic transformation for two modes}) contains no single-mode squeezing.
Consequently, $S_{\raisebox{0.0pt}{\tiny{A}}}$ is a pure two-mode squeezing operation,
$S_{\raisebox{0.0pt}{\tiny{A}}}=S_{\raisebox{0.0pt}{\tiny{TMS}}}(r)$, the paradigm Gaussian entangling operation,
where $r\in\mathbb{R}$ is the squeezing parameter. For initial states of two modes that are proportional to the identity,
as in our case, we can consider a \emph{resonance condition} as discussed in Ref.~\cite{BruschiDraganLeeFuentesLouko2012}, i.e.,
\begin{align}
    \comm{S}{S^{\,T}}   &=\,0\,.
    \label{eq:resonance condition}
\end{align}
Since the typical Bogoliubov transformations in analogue gravity systems do not contain any single mode squeezing, i.e.,
$S=S_{\raisebox{0.0pt}{\tiny{A}}}S_{\raisebox{0.0pt}{\tiny{TMS}}}$, the condition of Eq.~(\ref{eq:resonance condition})
has a very intuitive interpretation. It suggests that the state $\Gamma_{\!\raisebox{0.0pt}{\tiny{TMS}}}=S_{\raisebox{0.0pt}{\tiny{TMS}}}S_{\raisebox{0.0pt}{\tiny{TMS}}}^{\,T}$
 is invariant under the passive transformation $S_{\raisebox{0.0pt}{\tiny{P}}}$,
 $S_{\raisebox{0.0pt}{\tiny{P}}}\Gamma_{\!\raisebox{0.0pt}{\tiny{TMS}}}S_{\raisebox{0.0pt}{\tiny{P}}}^{\,T}=\Gamma_{\!\raisebox{0.0pt}{\tiny{TMS}}}$.
Since two-mode squeezing operations form a one parameter subgroup of the symplectic transformations,
$S_{\raisebox{0.0pt}{\tiny{TMS}}}(r_{1})S_{\raisebox{0.0pt}{\tiny{TMS}}}(r_{2})=S_{\raisebox{0.0pt}{\tiny{TMS}}}(r_{1}+r_{2})$,
one can easily see that a transformation which satisfies the resonance condition~(\ref{eq:resonance condition}) will
accumulate entanglement when repeated. In particular, the entanglement of formation of $\Gamma_{\!\raisebox{0.0pt}{\tiny{TMS}}}(r)$
is given by $h(e^{-2|r|})$, see Eqs.~(\ref{eq:EoF for symmetric Gaussian states}) and (\ref{eq:entropy measure}). The increase of
entanglement \emph{for particular modes} is then a matter of tuning the transformation at hand to fulfill the resonance condition~(\ref{eq:resonance condition}).

\section{Sudden quench of a BEC}\label{sec:sudden quench}

\begin{figure}[h!]
\hspace{2.5cm}\includegraphics[width=0.7\columnwidth]{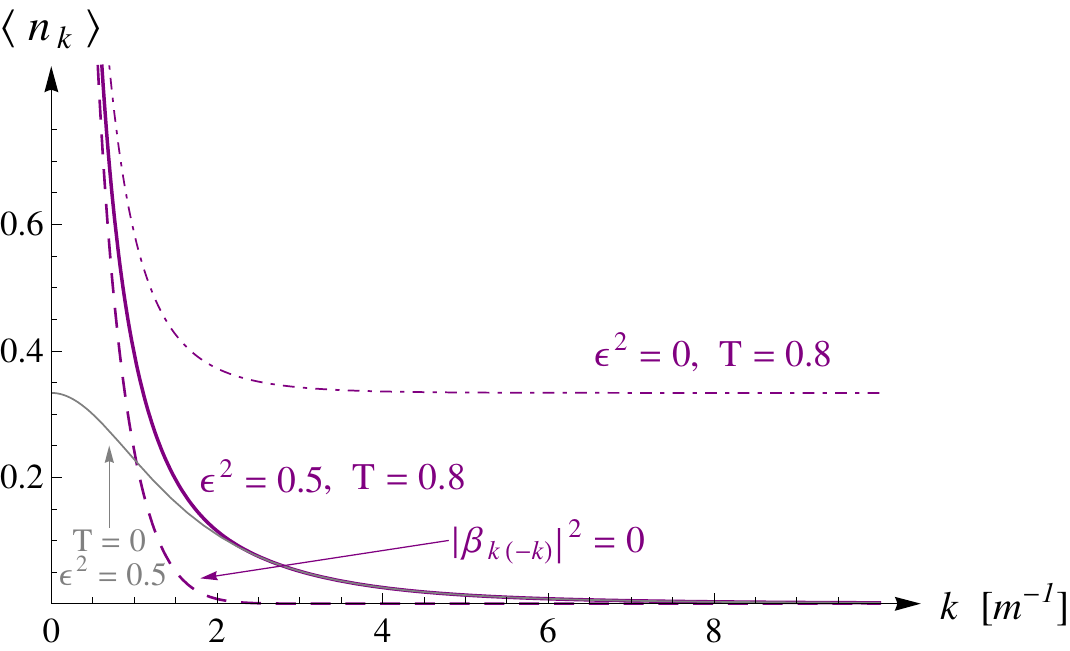}
\caption{Average particle number across the spectrum: The average particle number $\left\langle\,n_{k}\,\right\rangle$ after the sudden quench is plotted as a function of the wave number~$k$ for a nonlinear dispersion relation.
All curves are plotted for $c_{\,\mathrm{in}}=1$ms$^{-1}$ and $c_{\mathrm{out}}=3$ms$^{-1}$. Values for the parameter~$\epsilon$ are shown in units of $\mathrm{m}^{2}\mathrm{s}^{-1}$, while temperatures~$T$ are displayed in units of $(\hbar/k_{\mbox{\tiny{B}}})$K. For a thermal state at temperature~$T=0.5$ with nonlinear dispersion $(\epsilon^{2}=0.5)$ the initial average particle number (dashed purple curve) increases due to the sudden quench (solid purple curve). The increase is strongly suppressed for higher mode numbers with respect to the case of linear dispersion (dotted-dashed purple curve).
}
\label{fig:particle number plots}
\end{figure}

\begin{figure}[h!]
\hspace{2.5cm}\includegraphics[width=0.7\columnwidth]{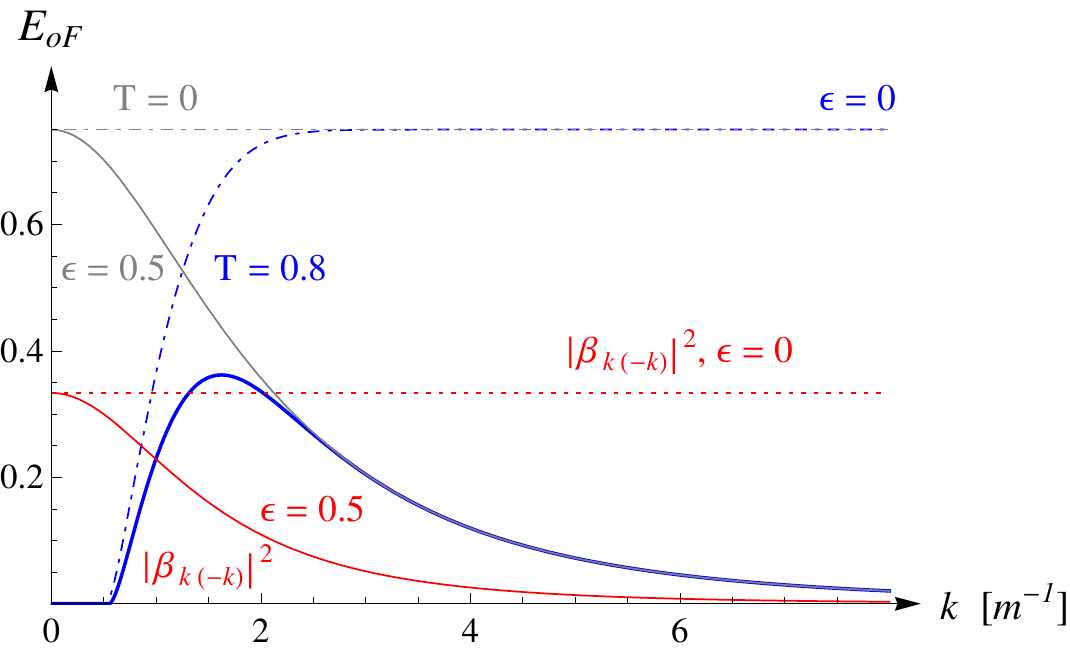}
\caption{Entanglement generation across the spectrum: The entanglement of formation $E_{oF}$ that is generated by a sudden quench is plotted as a function of the wave number $k$ with a nonlinear dispersion relation and nonzero temperature (solid thick blue line). For a linear dispersion relation $(\epsilon=0)$ and zero temperature, $T=0$, the generated entanglement (horizontal blue dashed line) and the particle production as measured by $|\beta_{k(-k)}|^{2}$ (horizontal red dotted line) are independent of~$k$. The nonlinearity of the dispersion relation, e.g., as shown here for $\varepsilon=0.5$ (in units of $\mathrm{m}^{2}\mathrm{s}^{-1}$), is damping the generation of particles (solid red curve [bottom]) and of the entanglement (solid gray line) for higher energy modes, i.e., large~$k$. Nonzero temperature, on the other hand, suppresses the entanglement generation for small energies (dotted-dashed blue line). All curves are plotted for $c_{\,\mathrm{in}}=1$ms$^{-1}$ and $c_{\mathrm{out}}=3$ms$^{-1}$. Temperatures are displayed in units of $(\hbar/k_{\mbox{\tiny{B}}})$K.}
\label{fig:quench plot}
\end{figure}

Let us now discuss a specific application of the general principles we have mentioned so far. We shall consider the Bogoliubov transformation that describes a single, sudden change of the speed of sound, a ``quench'', of a BEC at some initial temperature $T$. This can be achieved via a Feshbach resonance, i.e., a sudden change of the interaction strength, such that the density and phase of the BEC remain continuous, see Ref.~\cite{WeinfurtnerWhiteVisser2007} for details. As before we are going to make the approximation that the system is homogeneous throughout the process and that any effects of the inhomogeneity, in particular any caused by the quench, will enter as noise that reduces the entanglement.
The non-adiabatic adjustment of the speed of sound, $c_{\,\mathrm{in}}\rightarrow c_{\mathrm{out}}$,
causes the frequencies of the phononic modes to be altered, $\omega_{\,\mathrm{in}}\rightarrow\omega_{\mathrm{out}}$, while the
momenta $k$ remain the same. The Bogoliubov coefficients thus have exactly the previously discussed structure 
for $\Theta=0$. More specifically we have~\cite{WeinfurtnerWhiteVisser2007}
\begin{small}
\begin{subequations}
\label{eq:single quench bogos}
\begin{align}
    \alpha_{kk} &=\,\frac{1}{2}\Bigl(   \sqrt{\frac{\omega_{\mathrm{out}}}{\raisebox{1.0pt}{$\omega_{\mathrm{in}}$}}}\,+\,
                                        \sqrt{\frac{\omega_{\mathrm{in}}}{\raisebox{1.0pt}{$\omega_{\mathrm{out}}$}}}\,\Bigr)\,
    e^{i\,(\omega_{\mathrm{out}}\,-\,\omega_{\mathrm{in}})t_{0}},
    \label{eq:single quench alphas}\\[1mm]
    \beta_{k(-k)}   &=\,\frac{1}{2}\Bigl(\sqrt{\frac{\omega_{\mathrm{out}}}{\raisebox{1.0pt}{$\omega_{\mathrm{in}}$}}}\,-\,
                                        \sqrt{\frac{\omega_{\mathrm{in}}}{\raisebox{1.0pt}{$\omega_{\mathrm{out}}$}}}\,\Bigr)\,
    e^{-i\,(\omega_{\mathrm{out}}\,+\,\omega_{\mathrm{in}})t_{0}},
    \label{eq:single quench betas}
\end{align}
\end{subequations}
\end{small}
where $t_{0}$ is the time of the transition which becomes relevant for consecutive quenches. The frequencies are given by
the \emph{nonlinear dispersion relation} 
\begin{align}
\omega^{2}    &\,=\,c^{2}\,k^{2}\,\pm\,\epsilon^{2}\,k^{4}\,,
\label{eq:nonlinear dispersion}
\end{align}
with $\epsilon=\hbar/(2m)$, where $\hbar$ is Planck's constant, and $m$ is the mass of the atoms of the BEC. For a BEC only the positive sign \textemdash \emph{super-sonic} dispersion\textemdash occurs in Eq.~(\ref{eq:nonlinear dispersion}), but it is straightforward to consider the sub-sonic case for other systems. The nonlinear dispersion relation now enters the problem of entanglement creation due to the
Bogoliubov transformation in two places. First, the nonlinear effects influence the \mbox{$k$-dependence} of the initial
temperature distribution, i.e., the average number of phonons is still a thermal distribution but it is strongly
suppressed for higher mode numbers, as illustrated in Fig.~\ref{fig:particle number plots}. Additionally, the nonlinearity enters
directly in the Bogoliubov coefficients~(\ref{eq:single quench bogos}), which decreases the number of quasi-particles
that are produced by the quench. Together, the effects of the nonlinearity and the initial temperature compete to determine
the entanglement generation in Eq.~(\ref{eq:nu minus}). It thus becomes evident that, given a specific transformation and dispersion
relation, the entanglement generation is optimal for particular regions in $k$-space. We have illustrated this behaviour in
Fig.~\ref{fig:quench plot} for convenient, but not necessarily experimentally accessible, values of the parameters
$T$,~$\epsilon$,~$c_{\mathrm{in}}$ and~$c_{\mathrm{out}}$.





Finally, we can attempt to construct a resonant transformation from the single quench to enhance the entanglement production. For this transformation to
be repeatable it has to take $\omega_{\mathrm{in}}\rightarrow\omega_{\mathrm{in}}$. We can achieve a nontrivial transformation of this type by combining a
first quench, with $\omega_{\mathrm{in}}\rightarrow\omega_{\mathrm{out}}$, at time $t_{1}$ with a second quench that takes $\omega_{\mathrm{out}}\rightarrow\omega_{\mathrm{in}}$ at time $t_{2}$. We then evaluate the resonance condition~(\ref{eq:resonance condition}) for the
total transformation for which we find the two necessary conditions
\vspace*{-3mm}
\begin{subequations}
\label{eq:coupled quench resonance conditions}
\begin{align}
    \cos(\omega_{\mathrm{in}}t_{\mbox{\tiny{$+$}}})\,\sin(\omega_{\mathrm{out}}t_{\mbox{\tiny{$-$}}})\,f(\omega_{\mathrm{in}},\omega_{\mathrm{out}},t_{1},t_{2})  &=\,0\,,
    \label{eq:coupled quench resonance condition 1}\\[1mm]
    \sin(\omega_{\mathrm{in}}t_{\mbox{\tiny{$+$}}})\,\sin(\omega_{\mathrm{out}}t_{\mbox{\tiny{$-$}}})\,f(\omega_{\mathrm{in}},\omega_{\mathrm{out}},t_{1},t_{2})  &=\,0\,,
    \label{eq:coupled quench resonance condition 2}
\end{align}
\end{subequations}
where $t_{\mbox{\tiny{$\pm$}}}=(t_{1}\pm t_{2})\,$, and
\begin{align}
    f
    &=\,
    \omega_{\mbox{\tiny{$+$}}}^{2}\,\sin(\omega_{\mbox{\tiny{$-$}}}t_{\mbox{\tiny{$-$}}})\,-\,
    \omega_{\mbox{\tiny{$-$}}}^{2}\,\sin(\omega_{\mbox{\tiny{$+$}}}t_{\mbox{\tiny{$-$}}})\,,
\label{eq:remaining terms}
\end{align}
with $\omega_{\mbox{\tiny{$\pm$}}}=(\omega_{\mathrm{in}}\pm\omega_{\mathrm{out}})$. The equations~(\ref{eq:coupled quench resonance conditions}) are trivially
satisfied if $|t_{\mbox{\tiny{$-$}}}|=(n\pi/\omega_{\mathrm{out}})$ for any $n\in\mathbb{N}$. In these cases the combined transformation reduces to local
rotations that do not change the initial state. However, for $|t_{\mbox{\tiny{$-$}}}|\neq(n\pi/\omega_{\mathrm{out}})$ the remaining condition $f(\omega_{\mathrm{in}},\omega_{\mathrm{out}},t_{1},t_{2})=0$ can still suffice. Then the resonance condition reduces to the transcendent equation
\vspace*{-2mm}
\begin{align}
    \sin(\omega_{\mbox{\tiny{$-$}}}t_{\mbox{\tiny{$-$}}})/\omega_{\mbox{\tiny{$-$}}}^{2}   &=\,
    \sin(\omega_{\mbox{\tiny{$+$}}}t_{\mbox{\tiny{$-$}}})/\omega_{\mbox{\tiny{$+$}}}^{2}\,,
    \label{eq:transcendent equation}
\end{align}
which can be solved numerically. However, for some special values an analytical solution lies close at hand. For instance,
the transformation can be picked such that the ratio of $\omega_{\mbox{\tiny{$+$}}}$ and $\omega_{\mbox{\tiny{$-$}}}$ is rational, i.e., $m\omega_{\mbox{\tiny{$-$}}}=n\omega_{\mbox{\tiny{$+$}}}$, $m,n\in\mathbb{Z}$. Inserting this into (\ref{eq:transcendent equation}) the
transcendent equation can be easily solved for $t_{\mbox{\tiny{$-$}}}=n\pi/\omega_{\mbox{\tiny{$-$}}}$. Since we are excluding the trivial transformations
for which $t_{\mbox{\tiny{$-$}}}=(l\pi/\omega_{\mathrm{out}})$, $l\in\mathbb{Z}$, we obtain the resonant
solutions of Eq.~(\ref{eq:coupled quench resonance conditions}) by additionally requiring that $m$ and $n$ have odd separation, i.e., $(m-n)\neq2l$.

While such a series of sudden transformation might be only a rough estimation of the sinusoidal modification of the speed of sound applied in Ref.~\cite{JaskulaPartridgeBonneauRuaudelBoironWestbrook2012}, our framework allows to estimate whether a given system is above or below the sudden death temperature by measuring the average particle number for a given frequency and the initial temperature. With this information, Eq.~(\ref{eq:average particle number}) can be used to provide $T_{\mathrm{SD}}$ and determine if entanglement is present even if the explicit Bogoliubov coefficients are not known. Since tests of entanglement in analogue systems can be rather involved, this check is vital to ensure the viability of such experiments. As mentioned above our results are in principle applicable to both super- and sub-luminal types of non-linear dispersion, for instance surface waves.

\section*{Conclusions}

While our analysis assumes that the transformations of interest mix only couples of degrees of freedom (i.e., within homogeneous systems), the sudden death temperature we provide represents an upper bound on the allowed temperature for the inhomogeneous case as well. In particular in the case of resonant enhancement, e.g., by driving the transformation at a fixed frequency~\cite{JaskulaPartridgeBonneauRuaudelBoironWestbrook2012}, the inhomogeneity leads to a smearing of the sharp peaks and the entanglement is distributed over several adjacent modes.

In conclusion, we have conducted an analysis of the entanglement generation in analogue gravity systems at finite initial temperature. We find that the entanglement generation is fully determined by the Bogoliubov transformations describing the simulated gravitational, or relativistic effects. For every pair of quasi-particle modes of the system, the problem can be phrased in terms of an effective entanglement temperature $T_{\mathrm{E}}$.
If the initial temperature is above the benchmark of $2(\omega_{\mathrm{in}}/\raisebox{1.0pt}{$\omega_{\mathrm{out}}$})T_{\mbox{\tiny{E}}}$ then no entanglement is produced between the particular modes corresponding to $T_{\mbox{\tiny{E}}}\,$, regardless of the homogeneity of the system.
The detection of entanglement in analogue gravity systems is a major ambition of future setups, for instance, to test Bell inequalities~\cite{StobinskaJeongRalph2007} in similar settings as in Ref.~\cite{JaskulaPartridgeBonneauRuaudelBoironWestbrook2012}. Our results provide clear-cut criteria for the feasibility of such endeavours that are applicable to a broad range of current analogue gravity experiments. In particular we want to direct attention to Ref.~\cite{BuschParentani2013}, which appeared shortly after submission of our work and analyses closely related questions.

\ack
\vspace*{-5mm}
We thank Denis Boiron, Luis~J. Garay, Marcus Huber, Antony R.~Lee, Raphael Lopes, Carlos Sab\'{i}n, Chris Westbrook, Angela White, and Ivar Zapata for useful discussions and comments. S.~W. would like to thank Piyush Jain for useful comments and discussions.
N.~F. and I.~F. acknowledge support from EPSRC (CAF Grant No.~EP/G00496X/2 to I.~F.). D.~E.~B. acknowledges support from EPSRC (Grant No.~EP/J005762/1) and the European Community's Seventh Framework Programme under Grant Agreement No.~277110. N.~F. and D.~E.~B. also want to thank SISSA for hospitality and support.
S.~W. was supported by Marie Curie Actions \textemdash\ Career Integration Grant (CIG); Project acronym: MULTIQG-2011, the FQXi Mini-grant ``Physics without borders'', and the Royal Society University Research Fellowship ``Quantum Gravity Laboratory''.

\end{document}